\documentclass{jpp}

\usepackage{graphicx}
\usepackage{epstopdf, epsfig}
\usepackage{ulem}

\usepackage{amssymb}
\usepackage{amsfonts}
\usepackage{amsmath}
\usepackage{bm}
\usepackage{xcolor}
\usepackage{color}
\usepackage{amssymb}
\usepackage{natbib}
\usepackage{soul}

\shorttitle{ViDA: a Vlasov-DArwin solver for plasma physics at electron scales}
\shortauthor{O. Pezzi et al.}

\title{ViDA: a Vlasov-DArwin solver for plasma physics at electron scales}

\author{Oreste Pezzi\aff{1,2,3} \corresp{\email{oreste.pezzi@gssi.it}},
  Giulia Cozzani\aff{4,5}, Francesco Califano\aff{4}, Francesco Valentini\aff{3}, Massimiliano Guarrasi\aff{6}, Enrico Camporeale\aff{7,8}, Gianfranco Brunetti\aff{3}, Alessandro Retin\`o\aff{5} \and Pierluigi Veltri\aff{3}}

\affiliation{\aff{1} Gran Sasso Science Institute, Viale F. Crispi 7, I-67100 L’Aquila, Italy
\aff{2}INFN/Laboratori Nazionali del Gran Sasso, Via G. Acitelli 22, I-67100 Assergi (AQ), Italy
\aff{3} Dipartimento di Fisica, Universit\`a della Calabria, Via P. Bucci, I-87036 Arcavacata di Rende (CS), Italy
\aff{4} Dipartimento di Fisica ``E. Fermi'', Universit\`a di Pisa, Largo B. Pontecorvo 3, I-56127 Pisa, Italy
\aff{5} Laboratoire de Physique des Plasmas, CNRS/\'Ecole Polytechnique/Sorbonne Universit\'e, Universit\'e Paris Sud,
Observatoire de Paris, 91128 Palaiseau, France
\aff{6} CINECA Interuniversity Consortium, Via Magnanelli 6/3, 40033 Casalecchio di Reno, Italy
\aff{7} CIRES, University of Colorado, Boulder, CO, USA
\aff{8} Center for Mathematics and Computer Science (CWI), 1090 GB Amsterdam, The Netherlands }

\newcommand{\pg}[1]{{\color{orange}#1}} 
 
\begin{document}

\maketitle

\begin{abstract}
We present a Vlasov-DArwin numerical code (ViDA) specifically designed to address plasma physics problems, where small-scale high accuracy is requested even during the non linear regime to guarantee a clean description of the plasma dynamics at fine spatial scales. The algorithm provides a low-noise description of proton and electron kinetic dynamics, by splitting in time the multi-advection Vlasov equation in phase space. Maxwell equations for the  electric and magnetic fields are reorganized according to Darwin approximation to remove light waves. Several numerical tests show that ViDA successfully reproduces the propagation of linear and nonlinear waves and captures the physics of magnetic reconnection. We also discuss preliminary tests of the parallelization algorithm efficiency, performed at CINECA on the Marconi-KNL cluster. ViDA will allow to run Eulerian simulations of a non-relativistic fully-kinetic collisionless plasma and it is expected to provide relevant insights on important problems of plasma astrophysics such as, for instance, the development of the turbulent cascade at electron scales and the structure and dynamics of electron-scale magnetic reconnection, such as the electron diffusion region.
\end{abstract}

\section{Introduction}
\label{sect:intro}

Despite being studied with great efforts for about a century, natural and laboratory plasmas exhibit several complex phenomena that still need to be understood, mainly because of the strongly non-linear interactions and the presence of kinetic effects. In this context, investigating plasma dynamics is decisive for understanding fundamental processes occurring in different systems, ranging from very-far astrophysical objects to near-Earth environment and laboratory fusion devices. These systems routinely present a strongly nonlinear dynamics, which develops on a large range of spatial and time scales, including the ones associated with kinetic processes. In such systems, energy is typically injected at large fluid scales and cascades towards smaller scales, driving the system to cross three different physical regimes, ranging from fluid (MHD, Hall-MHD) to ion kinetic and eventually to electron kinetic scales. This multi-scale physics is the direct consequence of the weak plasma collisionality, that characterizes solar-wind and astrophysical plasmas~\citep{kulsrud2005plasma, califano2008one, bruno2016turbulence} as well as fusion devices dynamics, where collisions can become effective at scales smaller than the electron kinetic scales~\citep{falchetto2008european}.

As a result, the plasma is allowed to freely access the entire phase space and to manifest dynamical states far from thermal equilibrium~\citep{valentini2005self,galeotti2005asymptotic, marsch2006kinetic, franci2015solarwind, servidio2015kinetic, servidio2017magnetospheric, franci2018solarwind, sorriso2018local, pezzi2018velocityspace, cerri2018dual, sorriso2019turbulence}. As an example, we highlight here the fundamental role of the collisionless magnetic reconnection, that --even within a fluid theory framework-- drives a strongly nonlinear dynamics (at both ion and electron scales),  without collisions to be relevant~\citep{califano2007collisionless}. Within this context, the Vlasov equation for each particle species, self-consistently coupled to Maxwell equations for fields, provides a complete description of the system dynamics, although in some cases the role of weak collisions should be also  considered~\citep{navarro2016structure,pezzi2016collisional,pezzi2019protonproton}. The Vlasov-Maxwell model is an nonlinear integro-differential set of equations in multi-dimensional phase space, whose analytic solutions are only available in a few simplified cases and in reduced phase-space geometry. A numerical approach is therefore mandatory to describe the dynamics of collisionless magnetized plasmas in fully nonlinear regime.

As of today, numerical simulations have provided significant insights on the plasma dynamics at proton and sub-proton spatial scales, where proton kinetic effects are dominant, while electrons can be approximated as an isothermal fluid (hybrid framework) \citep{valentini2007hybrid}. In this range of scales, both Particle-In-Cell (PIC) and Eulerian hybrid codes have been extensively employed to investigate in detail a variety of physical phenomena such as, for instance, the development of the intermittent cascade of turbulent fluctuations~\citep{parashar2009kinetic, valentini2010twodimensional, servidio2012local, franci2015solarwind, servidio2015kinetic, franci2016plasma, valentini2016differential, cerri2017kinetic, cerri2017reconnection, valentini2017transition, pezzi2017revisiting, pezzi2017colliding, franci2018solarwind, perrone2018fluid, cerri2018dual, sorriso2018statistical}, the dynamo effect in turbulent plasmas \citep{rincon2016turbulent}, the interaction of solar wind and Earth's magnetosphere at global scales~\citep{palmroth2013preliminary, kempf2013wave, pokhotelov2013ion, vonalfthan2014vlasiator, hoilijoki2016mirror, pfaukempf2018importance, palmroth2018vlasov} and the dynamics of magnetic reconnection \citep{birn2001geospace, shay2001alfvenic, pritchett2008collisionless, califano2018electron}. To reduce the computational cost of the simulation reduced models --such as the gyro-kinetics \citep{howes2006astrophysical, howes2008kinetic, howes2008model, schekochihin2008gyrokinetic, tatsuno2009nonlinear, howes2011gyrokinetic, tenbarge2013collisionless, told2015multiscale, howes2016dynamical} or the finite Larmor radius Landau fluid ones \citep{passot2007collisionless, sulem2015landau, sulem2016influence, kobayashi2017three}-- have been also widely adopted to describe plasma dynamics at kinetic scales.

Within the context of space plasmas, recent high-resolution observations conducted by the Magnetospheric Multi-Scale (MMS) mission~\citep{burch2016magnetospheric, fuselier2016magnetospheric} allowed, for the first time, to investigate the plasma dynamics at electron scale. The MMS mission focuses primarily on kinetic processes occurring in the electron diffusion region of magnetic reconnection  ~\citep{burch2016electron, torbert2016estimates, torbert2018electronscale} and its unprecedented high resolution observations confirm a very complex picture where several mechanisms can be at work in producing small-scale fluctuations~\citep{lecontel2016whistler,breuillard2018new, chasapis2018insitu}. Magnetic reconnection often takes place within a turbulent environment where coherent structures --such as current sheets and X-points-- naturally develop~\citep{retino2007insitu, servidio2009magnetic, servidio2010statistics, haggerty2017exploring, phan2018electron}. At the same time, plasma jets generated by magnetic reconnection can provide energy for sustaining the turbulence itself~\citep{pucci2017properties,cerri2017kinetic,pucci2018generation}. Reconnection is important for space and astrophysical plasmas as it is responsible for major plasma heating and particle acceleration in solar and stellar coronae, magnetars, accretion disks and astrophysical jets~\citep{lyutikov2003explosive,uzdensky2011magnetic} as well as for tokamaks, being a major cause of loss of plasma confinement and plasma heating~\citep{helander2002runaway, tanabe2015electron}.  

In order to properly combine and compare the experimental evidences at electron scales with theoretical investigations~\citep{hesse2016theory}, a huge numerical effort needs to be done yet. To this end, only few numerical algorithms which retain both proton and electron kinetic physics are nowadays available. Most of them are PIC codes~\citep{markidis2010multi, daughton2011role, zeiler2002threedimensional, camporeale2011dissipation, karimabadi2013coherent, leonardis2013identification, divin2015evolution, wan2015intermittent, lapenta2015secondary, groselj2017fully, yang2017energyPOP, parashar2018dependence, shay2018turbulent, lapenta2019sonata, gonzalez2019turbulent}, which capture the full dynamics (including electron scales) since their computational cost is smaller with respect to low-noise Eulerian (Vlasov) codes. However, at variance with noise-free Eulerian algorithms, PIC codes fail in providing a clean description of small-scale fluctuations (e.g., the electric field behavior around the X-point) and particle distribution functions in phase space, since they suffer from intrinsic statistical noise. Only very recently, the first attempts to describe plasma dynamics {\it via} Eulerian fully-kinetic codes have became affordable, thanks to the improved supercomputer capabilities~\citep{schmitz2006darwin, umeda2009twodimensional, umeda2010full, tronci2015neutral, delzanno2015multi, umeda2016secondary, umeda2017nonMHD, ghizzo2017vlasov, juno2018discontinuous, roytershteyn2019numerical, skoutnev2019temperature}. As stated above, Eulerian algorithms generally require a computational cost significantly large as compared to PIC codes. A way to reduce the computational cost of a fully-kinetic Eulerian simulation consists in applying the so-called Darwin approximation \citep{kaufman1971darwin, birdsall2004plasma, schmitz2006darwin, schmitz2006kinetic} to the Maxwell equations based on the expansion of the Maxwell system in the small parameter $v^2/c^2$~\citep{mangeney2002numerical} ($v$ being the typical plasma bulk speed). Within this approximation, all wave modes (including those triggered by charge separation) are retained except for light waves ($v_\phi \sim c$, $v_\phi$ being the wave phase speed); by doing so, the numerical stability condition for the timestep can be significantly relaxed.

In the present work we present a newly developed fully-kinetic Eulerian Vlasov-DArwin algorithm (ViDA) which integrates numerically the kinetic equations for a non-relativistic globally-neutral plasma composed of protons and electrons. Equations are discretized on a fixed-in-time grid in phase space with periodic boundary conditions in the physical domain. ViDA originates from the hybrid Vlasov-Maxwell  code~\citep{valentini2007hybrid} (hereafter referred as HVM code) and has been extended specifically to include electron kinetic dynamics. The paper is organized as follows. In Sect. \ref{sect:darwin} we revisit the Darwin approximation and describe the system of equations, that is numerically integrated through ViDA. We discuss in detail the strategy of the numerical integration of the Vlasov equation for each species and we show that the Darwin version of the Maxwell equations can be written as a set of Helmholtz and Poisson-like equations, solvable through a spectral method. In the same Section, we also provide a description of the algorithm design. Then, in Sect. \ref{sect:numtests} we present first results obtained through this algorithm, concerning the propagation of i) electrostatic Langmuir waves, ii) whistler waves and iii) Alfv\'en waves. In Sect. \ref{sect:inst}, we describe the onset of the electron Weibel instability which is a plasma instability driven by the presence of a electron temperature anisotropy \citep{weibel1959spontaneously}. In Sect. \ref{sect:MR} we present preliminary results concerning one of the main potential applications of ViDA: the magnetic reconnection process at electron scales. Then, in Sect. \ref{sect:performance}, we discuss the performances of the algorithm. Finally, we conclude and summarize in Sect. \ref{sect:concl}.

\section{The Vlasov-Darwin (VD) model}
\label{sect:darwin}
The Darwin approximation, that we briefly revisit in the current section, has been adopted to reduce the limitations on the timesteps for numerical integration~\citep{kaufman1971darwin,mangeney2002numerical,birdsall2004plasma,schmitz2006darwin}. Indeed, since Maxwell equations allow for the propagation of waves at the light speed $c$, the timestep $\Delta t$ of any explicit numerical scheme solving these equations would be limited by the Courant-Friedrichs-Lewy (CFL) condition, $\Delta t \lesssim \Delta x/c$~\citep{peyret1986computational}. The Darwin approximation, by dropping the transverse displacement current term, rules out the transverse light waves (i.e. the fastest waves in the system that propagate at phase speed $c$) and significantly relaxes the CFL condition. 

We consider a non-relativistic, collisionless, fully-kinetic plasma composed by electrons and protons. The Vlasov-Darwin system of equations reads (in CGS units):
\begin{eqnarray}
&& \partial_t f_\alpha +  {\bm v} \cdot \nabla f_\alpha + \frac{Z_\alpha e}{m_\alpha} \left({\bm E} + \frac{{\bm v}}{c} \times {\bm B} \right) \cdot \nabla_{{\bm v}} f_\alpha = 0 \label{eq:vlas} \\
&& \nabla \cdot {\bm E}_L = 4 \pi \rho_c \label{eq:gaussE}\\
&& \nabla \cdot {\bm B} = 0 \label{eq:gaussB}\\ 
&&  \nabla \times {\bm E}_T = -\frac{1}{c} \partial_t {\bm B} \label{eq:faraday} \\
&&  \nabla \times {\bm B} = \frac{1}{c} \partial_t {\bm E}_L +\frac{4 \pi}{c} {\bm j} \label{eq:ampere}
\end{eqnarray}
where $f_{\alpha}({\bm x}, {\bm v}, t)$ is the distribution function (DF) of the $\alpha=p,e$ species, $m_\alpha$ and $Z_\alpha$ are respectively the mass and charge number of the $\alpha$ species and $c$ is the light speed. $\partial_t$, $\nabla$ and $\nabla_{{\bm v}}$ indicate the derivatives with respect to the time $t$, the spatial coordinates ${\bm x}$ and the velocity coordinates ${\bm v}$, respectively. ${\bm E}({\bm x},t)={\bm E}_L({\bm x},t) + {\bm E}_T({\bm x},t)$ and ${\bm B}({\bm x},t)$ are the electric and magnetic field, respectively. The electric field has been decomposed into a longitudinal (irrotational, $\nabla \times {\bm E}_L =0$) and a transverse (solenoidal, $\nabla \cdot {\bm E}_T = 0$) component \citep{griffiths1962introduction}. According to the Darwin approximation, the transverse component of the displacement current has been neglected in the Ampere's law [Eq. (\ref{eq:ampere})]~\citep{birdsall2004plasma, schmitz2006darwin}. The Darwin model, that retains at least the longitudinal component of the displacement current, is generally closer to the full Maxwell system with respect to models where the displacement current is completely neglected \citep{valentini2007hybrid,tronci2015neutral}.

The plasma charge density $\rho_c({\bm x},t)$ and the current density ${\bm j}({\bm x},t)$ are defined through the first two velocity moments of the particle DFs:
\begin{eqnarray}
&& \rho_c= e \sum_\alpha Z_\alpha n_\alpha = e \sum_\alpha Z_\alpha \int d\bm{v}\, f_\alpha \label{eq:n} \\
&& {\bm j} = \sum_\alpha\, {\bm j}_{\alpha} = e \sum_\alpha Z_\alpha n_\alpha {\bm V}_\alpha =  e \sum_\alpha Z_\alpha \int d\bm{v} \,
{\bm v} \, f_\alpha  \label{eq:j}
\end{eqnarray}

Equations (\ref{eq:vlas}--\ref{eq:ampere}) can be further simplified to obtain a set of Helmholtz-like equations of state without explicit time derivatives (see \citet{birdsall2004plasma,schmitz2006darwin} for details). By normalizing equations using a characteristic length $\bar{L}$, time $\bar{t}$, velocity $\bar{U}=\bar{L}/\bar{t}$, mass $\bar{m}$ and distribution function $f_{\alpha,0}=\bar{n}/\bar{U}^3$ (being $\bar{n}=\bar{L}^{-3}$ the equilibrium density), it is straightforward to get the dimensionless Vlasov-Darwin system of equations:

\begin{eqnarray}
&& \partial_t f_\alpha + \left( {\bm v} \cdot \nabla \right) f_\alpha + \frac{Z_\alpha}{\mu_\alpha } \left({\bm E} + {\bm v} \times {\bm B} \right) \cdot \nabla_{{\bm v}} f_\alpha = 0 \label{eq:vlasfin}  \\
&& \nabla^2 \varphi = - \zeta^2 \sum Z_\alpha n_\alpha \hspace{40pt} {\bm E}_L = - \nabla \phi \label{eq:poisEfin}\\
&& \nabla^2 {\bm B} = - \bar{u}^2 \zeta^2 \nabla \times {\bm j} \label{eq:poisBfin} \\
&& \nabla^2 \hat{{\bm E}}_T -\bar{u}^2 \zeta^2 \sum_\alpha \frac{Z_\alpha^2 n_{\alpha,0}}{\mu_\alpha} \hat{\bm E}_T = \bar{u}^2 \zeta^2 \left[ 
- \nabla \cdot \sum_\alpha Z_\alpha \langle {\bm v} {\bm v} \rangle_\alpha + \right. \nonumber \\
&& \left. +  \sum_\alpha \frac{Z_\alpha^2 }{\mu_\alpha} \left( n_\alpha {\bm E}_L + \langle {\bm v} \rangle_\alpha \times {\bm B} 
\right) \right]   \label{eq:ETfin} \\
&& \nabla^2 \Theta = \nabla \cdot \hat{{\bm E}}_T  \hspace{40pt} {{\bm E}}_T = \hat{{\bm E}}_T - \nabla \Theta  \label{eq:ETthfin} \\
&& \nabla \cdot {\bm B} = 0 \label{eq:gaussBfin}
\end{eqnarray}
where $\langle h \rangle_\alpha= \int d\bm{v} f_\alpha h$. In Eqs.  (\ref{eq:vlasfin}--\ref{eq:gaussBfin}), the electric and magnetic fields are normalized to $\bar{E}= \bar{m} \bar{U}/ e \bar{t}$ and $\bar{B} = \bar{m} c/ e \bar{t}$, respectively. Note also that we set $k_B=1$. Non-dimensional parameters are $\mu_\alpha = m_\alpha/\bar{m}$, $\bar{u} = \bar{U}/c$ and $\zeta=\bar{\omega}_p\bar{t}$, being $\bar{\omega}_p = \sqrt{4\pi e^2 \bar{n}/\bar{m}}$. Note that in Eq. (\ref{eq:ETfin}) we have omitted a term $~\bar{u}^2 \nabla \partial_{tt} \phi$ which could generate, in principle, an irrotational component, and we have introduced Eqs. (\ref{eq:ETthfin}) to preserve the solenoidality of ${\bm E}_T$ \citep{schmitz2006darwin}. The spatial dependence of $n_\alpha$ on the left-hand side of Eq. (\ref{eq:ETfin}) has been neglected ($n_\alpha\simeq n_{\alpha,0}$) to let coefficients be constant \citep{valentini2007hybrid}.

\subsection{Conservation properties}
It is straightforward to verify that Eqs. (\ref{eq:vlasfin}--\ref{eq:gaussBfin}) satisfy the mass $\int d{\bm x} d\bm{v} f_\alpha$ and entropy $S_\alpha = \int d\bm{x} d\bm{v} f_\alpha \log f_\alpha$ conservation. The energy conservation equation, obtained by multiplying Eq. (\ref{eq:vlasfin}) by $m_\alpha {\bm v}^2/2$, integrating over the phase-space volume $\int d\bm{x} d\bm{v}$ and summing over the species reads:
\begin{equation}
 \mathcal{E}_{kin} + \mathcal{E}_{th} + \mathcal{E}_{mag} + \mathcal{E}_{el} = const
 \label{eq:consener}
\end{equation}
where the kinetic energy is $\mathcal{E}_{kin} = \sum_\alpha (m_\alpha / 2) \int d\bm{x}\, n_\alpha {\bm u}_\alpha^2$, the thermal energy is $\mathcal{E}_{th} = \sum_\alpha (3 / 2) \int d\bm{x}\, n_\alpha T_\alpha$, the magnetic energy is $\mathcal{E}_{mag} = \sum_\alpha (\bar{m} / 2 \bar{u}^2 \zeta^2 ) \int d\bm{x}\, {\bm B}^2$ and the electrostatic energy is $\mathcal{E}_{el} = \sum_\alpha (\bar{m} / 2 \zeta^2 ) \int d\bm{x}\, {\bm E}_{L}^2$. Note that the temperature of the $\alpha$-species is defined as $3 n_\alpha T_\alpha/ m_\alpha = \int d\bm{v} ({\bm v} - {\bm u}_\alpha)^2 f_\alpha$ and, to get Eq. (\ref{eq:consener}), we have used $\int d\bm{x}\, {\bm w}_T \cdot {\bm w}_L = 0$, ${\bm w}_T$ and ${\bm w}_L$ being a generic transverse and longitudinal vector, respectively. In each of the tests described in the present work we have checked the conservation of these quantities: their variation with respect to initial values is always smaller than the $1\%$.

\subsection{ViDA algorithm and code design}
The Vlasov equation for each species is integrated numerically by employing the time splitting method first proposed by~\citet{cheng1976integration} in the electrostatic limit and later extended to the full electromagnetic case~\citep{mangeney2002numerical}. Darwin equations are solved through standard Fast Fourier Transform (FFT) algorithms.

In our case, the splitting algorithm for Eq. (\ref{eq:vlasfin}) reads as follows:
\begin{eqnarray}
&& \partial_t f_\alpha + \left( {\bm v} \cdot \nabla \right) f_\alpha = 0 \label{eq:physspadv}\\
&& \partial_t f_\alpha + \frac{Z_\alpha}{\mu_\alpha } \left({\bm E} + {\bm v} \times {\bm B} \right) \cdot \nabla_{{\bm v}} f_\alpha = 0 \label{eq:velspadv}
\end{eqnarray}
In the first equation ${\bm v}$ is a parameter, while in the second equation ${\bm x}$, ${\bm E}$ and ${\bm B}$ are parameters. At time $t$, the solutions of Eqs. (\ref{eq:physspadv}--\ref{eq:velspadv}) can be written as $\Lambda_{\bm x}(t) F_\alpha({\bm x},{\bm v})$ and $\Lambda_{\bm v} F_\alpha({\bm x},{\bm v})$, respectively. In last expressions, $\Lambda_{\bm x}$ and $\Lambda_{\bm v}$ are the advection operators in physical and velocity space, whose explicit definition, based on the third-order Van Leer scheme, can be found in \citet{mangeney2002numerical}; while $F_\alpha({\bm x},{\bm v})=f_\alpha({\bm x},{\bm v},t=0)$ is the initial condition. Note that $\Lambda_{\bm v}$ depends also on the particle species $\alpha$ and, for the sake of simplicity, we avoid to explicitly report such dependence.

The splitting scheme is a symplettic, second order accurate in time (see \cite{mangeney2002numerical} where the stability condition of the advection operator is also discussed) and the numerical solution at $t=t_N=N \Delta t$ is given by: 
\begin{equation}\label{eq:spl}
    f_\alpha({\bm x},{\bm v},t_N) = \left[\Lambda_{\bm x}\left(\frac{\Delta t}{2} \right) \Lambda_{\bm v}\left(\Delta t \right) \Lambda_{\bm x}\left(\frac{\Delta t}{2} \right) \right]^N f_\alpha({\bm x},{\bm v},0)
\end{equation}

At $t=0$, the distribution function $F_\alpha({\bm x},{\bm v})=f_\alpha({\bm x},{\bm v},t=0)$ is first advected in physical space by a half time-step, obtaining ${\tilde f}_\alpha({\bm x},{\bm v},\Delta t/2)$. Then, the following structure is executed:
\begin{enumerate}
 \item Computing the moments of ${\tilde f}_\alpha$ and evaluating the electromagnetic fields ${\bm E}_L$, ${\bm E}_T$ and ${\bm B}$, at $t=\Delta t/2$, through Eqs. (\ref{eq:poisEfin}--\ref{eq:gaussBfin});
 \item Performing a time-step advection in velocity space: ${\hat f}_\alpha= \Lambda_{\bm v}(\Delta t) {\tilde f}_\alpha$.
\item Performing a time-step advection in physical space: ${\tilde f}_\alpha = \Lambda_{\bm x} (\Delta t){\hat f}_\alpha$.
\end{enumerate}
This last structure is repeated in the algorithm, according to Eq. (\ref{eq:spl}), in order to get the evolved distribution function at any time instant.

The phase space domain is discretized as follows. The physical space $D_{\bm x}=[0,L_x]\times[0,L_y]\times[0,L_z]$ is discretized with $N_x\times N_y \times N_z$ gridpoints and periodic boundary conditions are used. The velocity space $D_{\bm v,\alpha}=[-v^{max}_{\alpha,x},v^{max}_{\alpha,x}]\times[v^{max}_{\alpha,y},v^{max}_{\alpha,y}]\times[-v^{max}_{\alpha,z},v^{max}_{\alpha,z}]$ is discretized by $(2N_{\alpha,v_x}+1)\times (2N_{\alpha,v_y}+1) \times (2N_{\alpha,v_z} + 1)$ grid points. Velocity-space boundary conditions impose $f_\alpha(|v_i| > v^{max}_{\alpha,i})=0$ ($i=x,y,z$). In order to ensure mass conservation, $v^{max}_{\alpha,i}$ is typically set to be a large multiple of the thermal speed  $v_{th,\alpha}=\sqrt{T_\alpha/m_\alpha}$. 

The ViDA algorithm has been designed in such a way that the user can select (i) different normalizations of the model equations, (ii) the possibility of setting motionless protons and (iii) different dimensionalities of the physical-space domain ($1D$, $2D$, or $3D$), the velocity-space domain being always three-dimensional ($3V$). Within ViDA spatial vectors always have three components and can be function of one, two or three spatial variables, depending on the physical-space dimensionality. Since Darwin equations are a set of Helmholtz-like equations, initial perturbations have to be introduced through the particle DFs (and their moments): this represents a difference with respect to standard codes where also magnetic perturbations can be introduced.
 A check on the solenoidality of ${\bm B}$ and ${\bm E}_T$ is also implemented at each time step.

\begin{figure*}
\begin{center}
\includegraphics[width=0.8\columnwidth]{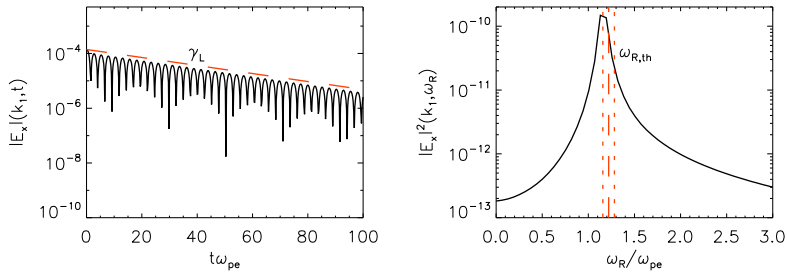}
\end{center}
\caption{Left: Time evolution of $|E_x|(k_1,t)$. The red dashed line indicates the predicted Landau damping rate $\gamma_L$. Right: Energy peak $|E_x|^2(k_1,\omega_R)$ as a function of the pulsation $\omega_R$. The red dashed line indicates the theoretical wave frequency $\omega_{R,th}$, while red dot-dashed lines show the $\omega_R$--resolution, i.e. $\omega_{R,th} \pm \Delta \omega_R/2$. The theoretical expectations for the Langmuir wave damping and the pulsation have been obtained with a numerical solver of the linear dispersion relation.}
\label{fig:LANlin}
\end{figure*}

The computational effort necessary to solve VD equations is significant and a massive parallelization, based on both MPI and OpenMP paradigms is implemented. The MPI paradigm, first introduced for the VDF by \citet{mangeney2002numerical}, is adopted to parallelize the physical-space computational domain for both particle DFs (and their moments) and electromagnetic field. Hence each MPI thread accesses a finite portion of phase space, composed by a sub-portion of physical space and by the whole velocity space. Within each MPI thread, the OpenMP directives are adopted to parallelize the velocity-space cycles. The parallelization of the electromagnetic field is a new feature recently introduced in the HVM code in \citet{cerri2017reconnection} and it is essential to perform high-resolution Eulerian Vlasov simulations, in particular in $3D$. Preliminary tests on performance and scalability are reported in Sect. \ref{sect:performance}.

\subsection{Normalizations of the Vlasov-Darwin equations}
\label{sect:normalizations}
In order to normalize Eqs.  (\ref{eq:vlasfin}--\ref{eq:gaussBfin}), three possible choices have been implemented in ViDA:
\begin{enumerate}
 \item {\it Electrostatic} normalization. Characteristic quantities are: length $\bar{L}=\lambda_{D,e}$, time $\bar{t}=\omega_{p,e}^{-1}$, velocity $\bar{U}=v_{th,e}$ and mass $\bar{m}=m_e$. Here $\lambda_{D,e}=\sqrt{T_e/4\pi n_e e^2}$, $\omega_{p,e}=(\sqrt{4\pi n_e e^2/m_e})^{-1}$, $v_{th,e}=\sqrt{T_e/m_e}=\lambda_{D,e}\omega_{p,e}$ and $m_e$ are the electron Debye length, the electron plasma frequency, the electron thermal speed and the electron mass, respectively. This normalization is appropriate for describing phenomena occurring at electron scales, such as the propagation of electrostatic plasma waves.
\item {\it Electromagnetic} normalization. Characteristic quantities are: length $\bar{L}=d_e$, time $\bar{t}=\omega_{p,e}^{-1}$, velocity $\bar{U}=c$ and mass $\bar{m}=m_e$, where $d_e=c/\omega_{p,e}$ is the electron skin depth. This normalization can be adopted for describing electromagnetic phenomena, where both protons and electrons are involved, such as magnetic reconnection and plasma turbulence at kinetic scales.
\item {\it Hybrid} normalization. Characteristic quantities are: length $\bar{L}=d_p$, time $\bar{t}=\Omega_{c,p}^{-1}$, velocity $\bar{U}=v_A$ and mass $\bar{m}=m_p$. In previous expressions $\Omega_{cp}=e B_0/m_p c$, $v_A=B_0/\sqrt{4\pi n_p m_p}$, $d_p=v_A/\Omega_{cp}$ and $m_p$ are the proton cyclotron frequency, the proton Alfv\'en speed, the proton skin depth and the proton mass, respectively. This normalization is useful for investigating the turbulent cascade in the sub-proton range, where electron physics starts to play a role.
\end{enumerate}
These three normalizations can be adopted to describe, in a more natural way (i.e. characteristic scales close to unity), phenomena where electrostatic, electromagnetic, or proton inertial effects dominate, respectively. 

\begin{figure*}
\begin{center}
\includegraphics[width=0.8\columnwidth]{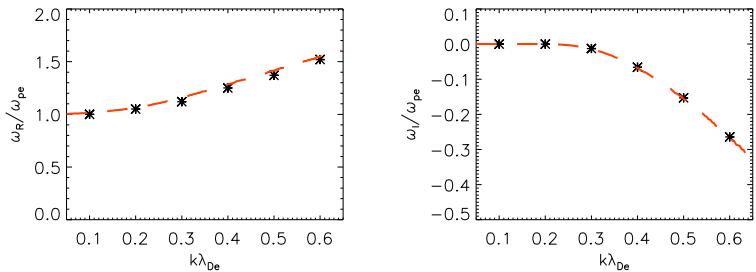}
\end{center}
\caption{Pulsation $\omega_R$ (left) and damping rate $\gamma_I$ (right), in units of $\omega_{p,e}$, from the simulation (black dots) and from the linear numerical solver (red dashed line) as a function of the wave number $k\lambda_{D,e}$.}
\label{fig:LANlinmanyk}
\end{figure*}

\section{Numerical tests of ViDA} \label{sect:numtests}

In this section we report the results of several tests performed to evaluate the capabilities of ViDA in describing basic collisionless plasma physics dynamics. The proper behavior and reliability of the code has been tested against the propagation of Langmuir waves, in both linear and nonlinear regime, whistler waves and Alfv\'en waves.

\subsection{Propagation and damping of Langmuir waves}
\label{sect:LAN}
For these tests we adopted the {\it electrostatic} normalization. We discuss results of simulations performed with motionless protons in $1D$--$3V$ phase-space configuration, where Langmuir waves propagate along the $x$ direction. Physical and velocity space have been discretized with $N_x=128$ and $N_{e,v_x}=50$, $N_{e,v_y}=N_{e,v_z}=15$ grid points, respectively. In the case of mobile protons ($m_p/m_e=1836$ and $T_p/T_e=1$), the propagation of Langmuir waves has been reproduced with lower phase space resolution, but the results are quantitatively similar to those discussed in the following. We have also separately tested the propagation of Langmuir waves along $y$ and $z$ directions by carrying out $2D$--$3V$ and $3D$--$3V$ runs.

\begin{figure*} 
\begin{center}
\includegraphics[width=1.0\columnwidth]{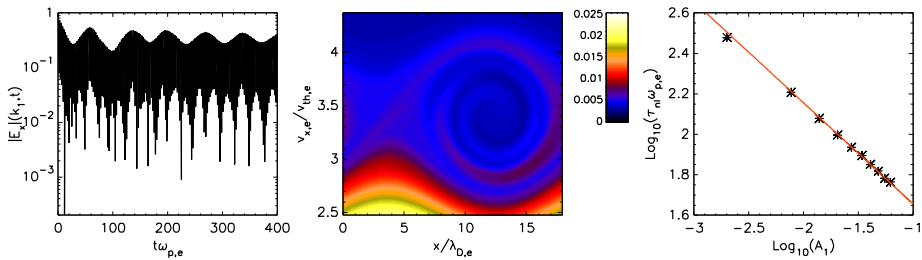}
\caption{ Left: Time evolution of $|E_{x}(k_1,t)|$ for the simulation with $A=8\times 10^{-2}$. Center: Contour plot of $\tilde{f_e}(x,v_{x})=\int dv_{y} dv_{z} f_e (x,{\bm v})$ in the plan $x$--$v_{x}$. Right: Nonlinear time $\tau_{nl}$ as a function of the first peak amplitude $A_1$. The red line reports the predicted scaling $\sim-0.5$ (the result of the linear fit is $\sim -0.48$). }
\label{fig:LANnonlin}
\end{center}
\end{figure*}

The initial equilibrium is given by an electron Maxwellian distribution spatially homogeneous. The plasma is unmagnetized, the initial electron temperature is $T_e=1$ (in scaled units). At $t=0$, the electron number density is perturbed through a sinusoidal perturbation $\delta n_e/n_{e,0} =A \sin(kx)$, $A=10^{-4}$ and $k=k_1=2\pi/L$  being the amplitude and the wave-number, respectively. The box length is $L=18\lambda_{D,e}$ ($k_1=0.35$) and $v^{max}_{e,i}= 5 v_{th,e}$ ($i=x,y,z$). The system evolution is reproduced up to a maximum time $t_{max}=100 \omega_{p,e}^{-1}$, while the numerical recurrence time is $t_{rec}=2\pi/k \Delta v \simeq 180 \omega_{p,e}^{-1}$\citep{cheng1976integration,galeotti2006echography, pezzi2016collisional}. 

The left panel of Fig. \ref{fig:LANlin} shows the time evolution of the amplitude of the  $k_x=k_1$ Fourier component of the electric field $|E_x|(k_1,t)$, in a semi-logarithmic plot. The electric field undergoes Landau damping \citep{landau1946vibration};  the observed damping rate shows a very good agreement with the theoretical prediction $\gamma_L=-3.37\times 10^{-2}\omega_{p,e}$ (red-dashed line), evaluated through a numerical linear solver for the roots of the electrostatic Vlasov dielectric function. In the right panel of Fig. \ref{fig:LANlin} we report the resonant curve, obtained by Fourier transforming the electric signal in space and time; we plot the spectral electric energy $|E_x|^2(k_1,\omega_R)$ as a function of the pulsation $\omega_R$. As expected, the resonant curve displays a well-defined frequency peak in correspondence of a value of the pulsation $\omega_R=1.22\omega_{p,e}$. In the right panel of the figure, the vertical red-dashed line represents the value of the theoretical resonant pulsation $\omega_{R,th}$ obtained through the linear solver, while the two vertical red-dot-dashed lines indicate the interval of uncertainty of the numerical code, due to the time discretization $\Delta \omega_R = 2\pi/t_{max} \simeq 0.063 \omega_{p,e}$. Again, numerical results are in very good agreement with theoretical predictions.

In order to show the dependence of real $\omega_R$ and imaginary $\omega_I$ part of the frequency as a function of the wavenumber, we have performed an additional $1D$--$3V$ run, in which the initial perturbation is a superposition of the first six wavenumbers $k_x=[k_1,6 k_1]$, where $k_1=2\pi/L$ ($L=2\pi 10 \lambda_{D,e}$); the other parameters are the same as in the previous run. To avoid numerical recurrence,  phase space has been discretized with $N_x = 128$, $N_{e,v_x}=100$ and $N_{e,v_y}=N_{e,v_z}=15$. Figure  \ref{fig:LANlinmanyk} reports by stars the dependence of $\omega_R$ (left) and $\omega_I$ (right), in units of $\omega_{pe}$, as  a function of the wave number $k\lambda_{D,e}$. A very good agreement with theoretical expectations (red-dashed curves) is recovered for both real and imaginary parts of the complex frequency. 

We conclude this section by focusing on the nonlinear regime of the Langmuir wave dynamics (see, for example, \citet{brunetti2000asymptotic} and refs. therein). We have performed ten different runs, varying the amplitude of the initial density perturbation in the range $A=[8\times10^{-3}, 8\times10^{-2}]$. In this case, phase space has been discretized with $N_x = 128$, $N_{e,v_x}=150$ and $N_{e,v_y}=N_{e,v_z}=15$, while $t_{max} = 400 \omega_{pe}^{-1}$. The box length is $L_x = 18 \lambda_{D,e}$, while $v^{max}_{e,i}= 5 v_{th,e}$ ($i=x,y,z$). As reported in the left panel of Fig. \ref{fig:LANnonlin}, the time evolution of the electric-field Fourier component shows an early linear damping phase~\citep{landau1946vibration}, until particle trapping arrests the damping and produces oscillations of the signal envelope~\citep{oneil1965collisionless}. At large times, a phase-space vortex is observed in the electron DF in the vicinity of the wave phase speed, as reported in the center panel of Fig. \ref{fig:LANnonlin}(center). As it has been shown in \citet{oneil1965collisionless}, the nonlinear trapping time $\tau_{nl}$ depends on the saturation amplitude $A_1$ of the electric oscillations as $\tau_{nl} \sim A_1^{-1/2}$. For each of the ten simulations, we evaluated $A_1$ and $\tau_{nl}$ at the time of the first peak of the electric envelope oscillations. The right panel of Fig. \ref{fig:LANnonlin} shows in log-log plot $\tau_{nl}$ as a function of $A_1$ (stars), compared to the theoretical expectation (red line), showing a very nice agreement.

\begin{figure*}
\begin{center}
\includegraphics[width=0.8\columnwidth]{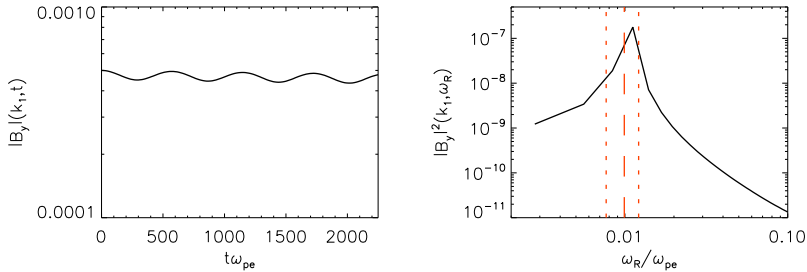}
\end{center}
\caption{Left: Time evolution of $|B_y|(k_1,t)$. Right: Spectral magnetic energy $|B_y|^2(k_1,\omega_R)$ as a function of the pulsation $\omega_R$. The red dashed line indicates the theoretical wave frequency $=\omega_{R,th}$, while red dot-dashed lines show the $\omega_R$-resolution $\omega_{R,th} \pm \Delta \omega_R/2$.}
\label{fig:WHlin}
\end{figure*}

\subsection{Propagation of whistler waves}
\label{sect:WH}
To reproduce the propagation of whistler waves at electron scales, the  {\it electromagnetic} normalization has been employed. Protons are assumed just as a fixed neutralizing background. Again, we have verified that the ViDA code behaves exactly in the same manner in the three spatial directions. Hence, we discuss here the result of a $1D$--$3V$ run, where ${\bm B_0}= B_0{\bm e}_x$ ($B_0=1$) and protons are not fixed. The box length is $L_x= 2\pi 10 d_e$, while $v^{max}_{e,i}= 10 v_{th,e}$ and $v^{max}_{p,i}= 7 v_{th,p}$ ($i=x,y,z$). Note that increasing the value of $v^{max}_{p(e)}$ has been necessary to ensure mass conservation. The phase space has been discretized with $N_x = 128$, $N_{e,v_x}=N_{e,v_y}=N_{e,v_z}=50$ and $N_{p,v_x}=N_{p,v_y}=N_{p,v_z}=35$. We also set $m_p/m_e = 1836$, $T_e/T_p=1$, $\bar{u}=v_{th,e}/c= 10^{-3}$ and $\zeta=1$. 

The equilibrium is composed of Maxwellian velocity distributions for both protons and electrons and homogeneous density. The initial equilibrium is then perturbed with the following electron bulk-speed perturbations:
\begin{eqnarray}
 && \delta u_{e,y} = A \sin(k x) \\
 && \delta u_{e,z} = A \cos(k x)
\end{eqnarray}
where $A=10^{-3}$ and $k=k_1=2\pi/L_x$. 

By solving Darwin equations, these perturbations generate a current density and then magnetic fluctuations. Figure \ref{fig:WHlin} reports the time evolution of $|B_y|(k_1,t)$ (left) and the frequency peak of the spectral magnetic energy $|B_y|^2(k_1,\omega_R)$ as a function of the pulsation $\omega_R$ (right). The magnetic field clearly oscillates at the correct frequency $\omega_{R,th}=0.91 \omega_{p,e}$, that can be evaluated from the linear dispersion relation for whistler waves (obtained by assuming motionless protons and cold electrons): $\omega_{R,th}(k) = B_0 k^2/(1 + k^2)$  \citep{krall1973principles}.
Note that, for the considered wave-number, a negligible damping of whistler waves is expected. In the left panel of Fig. \ref{fig:WHlin}, red-dashed line represents the value of the resonant pulsation from above expression for $\omega_{R,th}$, while the two vertical red dot-dashed lines indicate the interval of numerical uncertainty $\Delta \omega_R = 2\pi/t_{max} \simeq 0.063 \omega_{p,e}$.

\begin{figure*}
\begin{center}
\includegraphics[width=0.8\columnwidth]{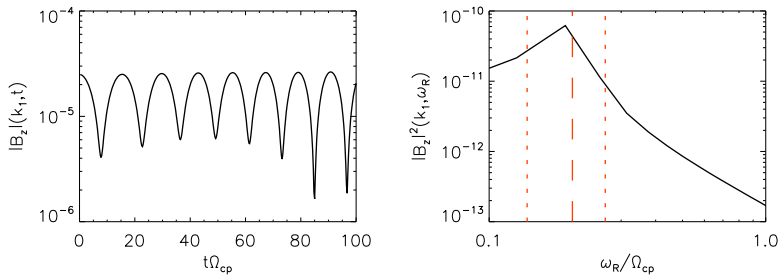}
\end{center}
\caption{Left: Time evolution of $|B_z|(k_1,t)$. Right: Magnetic spectral energy $|B_z|^2(k_1,\omega_R)$ as a function of $\omega_R$. The red solid line indicates the theoretical wave frequency $\omega_{R,th}$, while red dashed lines show the $\omega_R$-resolution $\omega_{R,th} \pm \Delta \omega_R/2$.}
\label{fig:AWxlin}
\end{figure*}

\subsection{Propagation of Alfv\'en waves}
\label{sect:AW}
Here we show numerical results concerning  the propagation of Alfv\'en waves along a background magnetic field. The adopted normalization for these tests is the {\it hybrid} one. We perform a $1D$--$3V$ run, where ${\bm B_0}= B_0{\bm e}_x$ and $B_0=1$. The box length is $L_x= 2\pi 50 d_p$, while $v^{max}_{e,i}= 5 v_{th,e}$ and $v^{max}_{p,i}= 5 v_{th,p}$ ($i=x,y,z$). The phase space has been discretized with $N_x = 32$, $N_{e,v_x}=N_{e,v_y}=N_{e,v_z}=25$, $N_{p,v_x}=40$, and $N_{p,v_y}=N_{p,v_z}=35$ gridpoints. The mass ratio has been artificially set $m_p/m_e = 25$, thus avoiding extremely small timesteps, while $T_e/T_p=1$, $\bar{u}=v_{A}/c= 10^{-3}$ and $\zeta=c/v_{A}=10^3$. Initial proton $\beta_p$ is $\beta_p=2 v_{th,p}^2/v_A^2 = 1$.

The initial equilibrium, composed of spatially homogeneous Maxwellian protons and electrons, has been perturbed with the following proton bulk-speed perturbations:
\begin{eqnarray}
 && \delta u_{p,y} = A \sin(k x) \\
 && \delta u_{p,z} = A \cos(k x)
\end{eqnarray}
where $A=10^{-4}$ and $k=k_1=2\pi/L_x$. Figure \ref{fig:AWxlin} shows the time evolution of $|B_z|(k_1,t)$ (left) and the magnetic spectral energy $|B_z|^2(k_1,\omega_R)$ as a function of $\omega_R$ (right). The recovered resonant peak is in agreement with the theoretical pulsation, evaluated through a fully-kinetic linear solver of the dispersion relation \citep{camporeale2017comparison}. Moreover, no Landau damping is observed, since it occurs at much smaller scales \citep{barnes1966collisionless, vasconez2014vlasov,camporeale2017comparison}.  This test represents the first attempt towards a general description of Alfv\'en waves, where electron physics is also taken into account. Since including electron physics is currently too computational demanding, we plan to continue the investigation in a separate, future work.

\begin{figure}
\centering
\includegraphics[width=0.6\columnwidth]{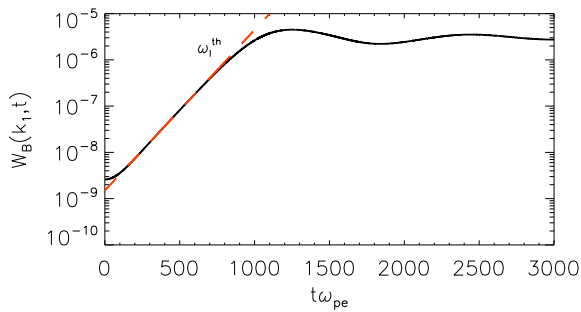}
\caption{Time evolution of $W_B(k_1)= (|B_y|^2(k_1) + |B_z|^2(k_1))/2$. The red dashed line indicates the linear instability growth $\omega_I^{th}\simeq 4\times 10^{-3}\omega_{p,e}$, calculated with a linear solver of the fully-kinetic dispersion relation \citep{camporeale2017comparison}. }
\label{fig:eWeib}
\end{figure}

\section{Temperature anisotropy driven instability: electron Weibel instability}
\label{sect:inst}

Another class of interesting numerical tests, which can be performed to point out the reliability of the ViDA code, concerns the onset of micro-instabilities, such as whistler, mirror and Weibel instabilities driven by a temperature anisotropy \citep{weibel1959spontaneously,gary2006linear,califano2008nonlinear,palodhi2009nonlinear, palodhi2010transition,chen2014energy,chen2015multi,camporeale2015wave}. \\

Here we focus on the development of the electron Weibel instability, that produces electromagnetic fluctuations transverse to the wavevector ${\bm k}$. The most suitable normalization to perform this analysis is the {\it electromagnetic} one. In particular we discuss the results of a 1D--3V run with ${\bm k}=k \hat{{\bm e}}_x$, although we have verified that instability is triggered in the same way also in the different phase space configuration ($2D$ and $3D$). The mass ratio is $m_p/m_e=100$, while $T_e/T_p=0.01$. Electrons are initialized with a bi-Maxwellian distribution function, with thermal speeds $v_{th,e,x}=2.5\times 10^{-2}c$ and $v_{th,e,y}=v_{th,e,z}=4 \times 10^{-2}c$, this giving a temperature anisotropy  $A=T_{y(z)}/T_x=2.56$. Protons have a Maxwellian velocity distribution at $t=0$, with a thermal speed $v_{th,p} = v_{th,e,x}$ and homogeneous density. However, in this case, protons mainly act just as a neutralizing background, not being involved in the dynamics during the linear stage (i.e. during the instability development). No background magnetic field has been introduced. Physical space, whose length is $L_x=32 d_e$, has been discretized with $N_x=64$ gridpoints. Velocity space has been discretized with $51^3$ gridpoints for both protons and electrons and  $v^{max}_{e(p)}=5v_{th,e(p)}$ in each velocity directions.

The initial equilibrium has been perturbed through a sinusoidal, transverse perturbation, imposed on the electron bulk speed:
\begin{eqnarray}
 && \delta u_{e,y} = A sin(k x) \\
 && \delta u_{e,z} = A cos(k x)
\end{eqnarray}
where $A=2\times10^{-5}$ and $k=k_1=2\pi/L_x$. Such bulk-speed perturbations produce a current density, which in turn generates magnetic fluctuations. Figure \ref{fig:eWeib} reports the time evolution of the magnetic spectral energy density $W_B(k_1,t)= (|B_y|^2(k_1,t) + |B_z|^2(k_1,t))/2$. The red-dashed line indicates the expected linear instability growth rate $\omega_I^{th}\simeq 4\times  10^{-3}\omega_{p,e}$, evaluated through a linear solver for the roots of the kinetic electromagnetic dielectric function. In the early stage of the simulation, $W_B$ increases exponentially with a growth rate in very good agreement with the expected one. Then, oscillations saturate at a nearly constant value in the nonlinear regime of evolution \citep{chen2014energy}.

\section{Dynamics of magnetic reconnection} 
\label{sect:MR}
In this Section we present the results of a magnetic reconnection simulation. Generally speaking, Vlasov simulations of magnetic reconnection represent a strong numerical challenge because of the huge memory and CPU time required by Eulerian algorithms. This approach, if successful, would certainly provide a crucial contribution to the understanding of the magnetic reconnection process especially at electron scales, thanks to the fact that Eulerian algorithms allow for an almost noise-free description of fields and particle DFs. A noise-free description is crucial to properly understand e.g. which electromagnetic fluctuations contribute to the reconnection electric field in the form of anomalous resistivity and how  distribution functions are modified leading to electron heating. 

We have performed a $2D$--$3V$ symmetric magnetic reconnection simulation. Reconnection is symmetric when the values of magnetic field and density are equal on the two opposite sides of the current sheet. The initial condition of our simulation is the one adopted in the GEM challenge \citep{birn2001geospace}, in order to allow for a direct comparison to previous studies \citep{birn2001geospace, schmitz2006kinetic}. For this reason, we have also chosen the  {\it hybrid} normalization (see Sect. \ref{sect:normalizations}).

The equilibrium is set by adapting the Harris equilibrium \citep{harris1962plasma} to the periodic boundary conditions in the spatial domain. In particular, the component of the magnetic field $B_x(y)$  corresponding to the double current sheet profile reads:
\begin{equation}
B_x(y) = B_{0}  \biggl[\tanh\biggl(\frac{y - L_y/2}{L_1}\biggr) - \tanh\biggl(\frac{y}{L_2}\biggr)  -  \tanh\biggl(\frac{y - L_y}{L_2}\biggr) \biggr].
\label{eq:bx}
\end{equation} 

This profile is characterized by the presence of two gradients (the current sheets) varying as an hyperbolic tangent and located at $y = L_y/2$ and $y = 0$ (and so at $y = L_y$) where $L_y$ is the length of the spatial domain in the $y$ direction.  
The first hyperbolic tangent is the one defined in \cite{harris1962plasma} and $L_1$ is the corresponding current sheet thickness. The second and third hyperbolic tangent in Eq.(\ref{eq:bx}) have been included to satisfy the spatial periodicity; the value of $L_2$ is taken sufficiently large compared to $L_1$ to slow down the development of reconnection in the second current sheet with respect to the main one.
The electron and ion temperature are set uniform at the initial time and the density $n(y)$ is defined in order to satisfy pressure balance. Then, from Eq. (\ref{eq:ampere}) and considering  $\partial_t {\bf E}_L = 0 $ at the initial time, we get the initial current density ${\bf j} = (0, \ 0, \ j_z(y))$.
 
 Following the prescriptions of the Harris equilibrium we get, in normalized units,
\begin{eqnarray}
n_0(T_e + T_p) = \frac{B_0^2}{2} \\
\frac{u_{e,0}}{T_e} = - \frac{u_{p,0}}{T_p} \label{eq:harris_v_condition}\\
\frac{j_z(y)}{n(y)} \equiv u_0 = u_{p,0} - u_{e,0} \label{eq:u0}
\end{eqnarray}
Eq. (\ref{eq:harris_v_condition}) corresponds to the no charge separation condition of the Harris equilibrium so that quasi-neutrality is imposed, $n_e(y) = n_p(y) = n(y)$. In other words, the electric field is zero at the initial time. Moreover, from Eqs.(\ref{eq:harris_v_condition})--(\ref{eq:u0}) we have:
\begin{eqnarray}
u_{e,0} = - \frac{u_{0}}{1+ T_p / T_e} \\
u_{p,0} = \frac{u_{0}}{1+ T_e / T_p} 
\end{eqnarray}
It is worth to point out that this is not an exact Vlasov kinetic equilibrium. In particular, it differs from the equilibrium presented by Harris since in this simulation the spatial domain is periodic in the varying $y$-direction. On the other hand, the initial configuration is in force balance and we have checked that the initial equilibrium is not significantly affected by, for example, ballistic effects within the time scale of reconnection considered here. 
As for the GEM challenge \citep{birn2001geospace}, fluctuations are superposed to the initial magnetic field in order to obtain a single magnetic island at the center of the space domain at the initial time. In particular, $\delta \mathbf{B} = \nabla \delta \psi \times \hat{z}$ and 
\begin{equation}
\delta \psi(x,y) = \psi_{0}  \cos(2 \pi x/L_x) \cos(2 \pi y/L_y)
\end{equation} 
where, as already stated, $L_x$ and $L_y$ are the lengths of the spatial domain in $x$ and $y$ direction respectively. According to GEM challenge, in scaled units, $\psi_0$ is set to $0.1$. 

\begin{figure*}
\begin{center}
\includegraphics[width=0.6\columnwidth]{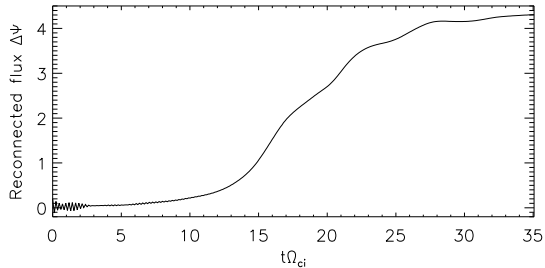}
\end{center}
\caption{Time evolution of the reconnected magnetic flux $\Delta \psi$.}
\label{fig:rec_flux}
\end{figure*}

By using the relation $\delta \mathbf{B}(x,y) = \nabla  \delta \psi(x,y) \times \hat{z}$ and Eq.(\ref{eq:poisBfin}), we derive the expression for the current density fluctuations $\delta \mathbf{j}(x,y)$ consistent with $\delta \psi(x,y)$. In particular, it is possible to define $\delta \mathbf{j}(x,y) = (0, 0, \delta j_z(x,y))$.
Finally, the initial electron and proton distribution functions are shifted Maxwellian distributions with drift velocities along the $z$ direction and temperature $T_e$ and $T_p$.

The phase space has been discretized with $N_x\times N_y = 512\times512$ gridpoints in the spatial domain,  $N_{e,v_x}\times N_{e,v_y}\times N_{e,v_z} =  41\times41\times81$ gridpoints in the velocity domain for electrons and $N_{p,v_x}\times N_{p,v_y}\times N_{p,v_z} = 31\times31\times31$ gridpoints in the velocity domain for protons. We also set $v_{e}^{max} = 5 \ v_{th,e}$ and $v_{p}^{max} = 5 \ v_{th,p}$, where the normalized $v_{th,p}$ is set to $1$. 
Other simulation parameters are $L_1 = 0.5 d_p$, $L_2 = 2.5 d_p$, $m_p/m_e = 25$, $n_{\infty} = 0.2$, $T_e/T_p = 0.2$, $L_x = L_y = 25.6 d_p$.  Also, we set $B_0 = 1$ and $n_0 = 1$. All parameters are chosen to be as close as possible to the simulation parameters listed in \cite{birn2001geospace}.

In Figure \ref{fig:rec_flux} we show the evolution of the reconnected flux given by the difference $\Delta \psi$ between the magnetic flux $\psi$ evaluated at the X point and at the O point.  Accordingly to the initial perturbation, the X and the O point are initially located at $(L_x/2, L_y/2)$ and $(0, L_y/2)$ and their location does not significantly change throughout the simulation run. The behavior of $\Delta \psi$ is very similar to the evolution of the reconnected flux in Ref. \citep{birn2001geospace}. Reconnection evolves with a reconnected flux that remains close to zero until $t \sim 15 \ \Omega_{c,p}^{-1}$, when a sharp increase is observed. Then, the reconnection rate stays relatively constant until the reconnected flux begins to saturate at $t \sim 30 \ \Omega_{c,p}^{-1}$. 

\begin{figure*}
	\begin{center}
	\includegraphics[width=\columnwidth]{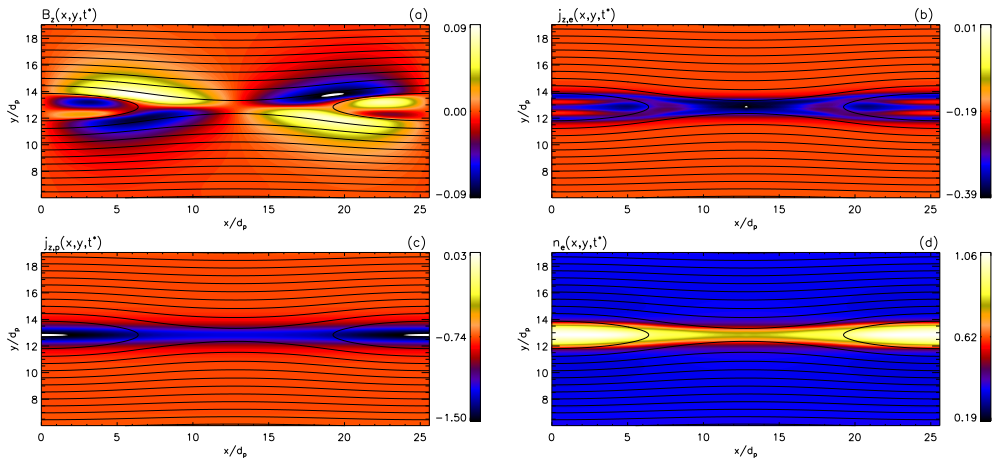}
	\end{center}
	\caption{Contour plots of $B_z$ (a); out-of-plane electron current density $j_{e,z}$ (b); out-of-plane proton current density $j_{p,z}$ (c); and electron number density $n_e$ (d). The quantities are shown at the time $t^* = 15.27 \ \Omega_{c,p}^{-1}$. At that time $\Delta \psi = 1.18$. All the panels are zoomed in $y$ in the interval $[6 \ d_p, 19 \ d_p]$.}
	\label{fig:bzjz}
\end{figure*}

In Figure \ref{fig:bzjz} we show the contour plots of the out of plane magnetic field $B_z$ (a), of the electron current density in the $z$-direction $j_{e,z}$ (b), of the proton current density in the $z$-direction $j_{p,z}$ (c) and of the electron number density $n_e$ (d). In each panel, the contour lines of the magnetic flux $\psi$ are superposed. $B_z$ exhibits the typical Hall quadrupolar pattern usually observed during symmetric magnetic reconnection. This magnetic signature indicates that the ions are demagnetized while the electrons are still frozen to the magnetic field. The difference in their dynamics produces the out-of-plane $B_z$ \citep{mandt1994transition, uzdensky2006physical}. The quadrupolar structure that we find is analogous to the one obtained with other kinetic codes, both Eulerian \citep[see Fig. 2]{schmitz2006kinetic} and Lagrangian \citep[see Plate 1(b)]{Pritchett:2001JGR}. We note that the $j_{p,z}$ pattern closely follows the density pattern ($n_e \simeq n_p$) so that $j_{p,z}$ is depleted at the X point while it reaches its maximum value within the magnetic island. On the other hand, $j_{e,z}$ is enhanced at the X point and the region of strong $j_{e,z}$ is elongated along $x$. Away from the X point, $j_{e,z}$ splits into two branches that identify the separatrices, as it was also observed by \citet{shay2001alfvenic} (see Fig.6(d)). The electron current at the X line has a thickness comparable to $d_p$ which corresponds to $5 \ d_e$. The maximum value of the normalized $B_z$ is $0.09$ while the maximum values of $j_{p,z}$ and $j_{e,z}$ are $0.39$ and $1.49$, respectively. These values are overall slightly smaller than the values found in a similar Vlasov-Darwin simulation described in Ref. \citep{schmitz2006kinetic}. 

\begin{figure*}
	\begin{center}
		\includegraphics[width=0.8\columnwidth]{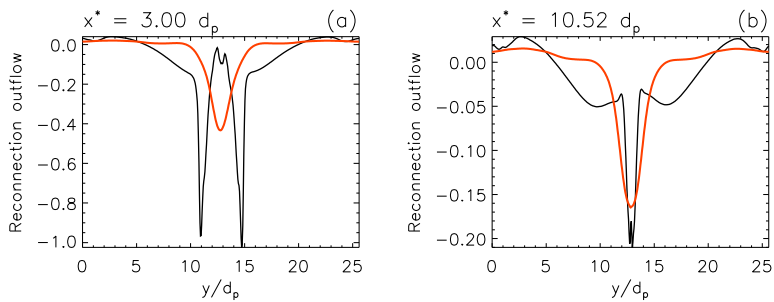}
	\end{center}
	\caption{(a) $x$ component of the electron velocity $u_e$ (black line) and of the proton velocity $u_p$ (red line) at $x^* = 3.00 \ d_p$ and (b) at $x^* = 10.52 \ d_p$. The quantities are shown at the time $t^* = 18.13 \ \Omega_{c,p}^{-1}$.}
	\label{fig:jet}
\end{figure*}

In Figure \ref{fig:jet} we show the reconnection outflow of protons and electrons at $t^* = 18.13 \ \Omega_{c,p}$. In particular, we note that at $x = 3 \ d_p$ (panel (a)), corresponding to a distance of $9.8 \ d_p$ from the X-point located at $L_x/2 = 12.8 \ d_p$, the electron velocity is characterized by two peaks corresponding to the separatrices, while the proton velocity is concentrated in the center of the outflow region and it reaches lower values, as expected. The presence of the two peaks is consistent with the $j_{e,z}$ pattern shown in Fig. \ref{fig:bzjz}(b).
Fig. \ref{fig:jet}(b) shows the same quantities of Fig.\ref{fig:jet}(a) at a distance of  $2.3 \ d_p$ from the X-point where the outflow is still developing and we note that $u_{e,x}$ is rather similar in shape and value to $u_{p,x}$.

\section{Performance test on the ViDA code}
\label{sect:performance}

In this section, we present preliminary performance tests of ViDA implemented on the Marconi-KNL cluster at the CINECA supercomputing center (Casalecchio di Reno (BO), Italy). The Marconi cluster is equipped with 3600 Lenovo Adam Pass nodes, interconnected through the Intel OmniPath network and each one composed by 1KNL processor (68 cores, 1.40GHz), formally 96 GB of RAM (effective 83 GB) and 16 GB of MCDRAM. The tests have been performed on a simple equilibrium configuration (Maxwellian DFs with no perturbations). We remark however that this choice does not affect the code performance. 

\begin{figure*}
	\begin{center}
		\includegraphics[width=0.6\columnwidth]{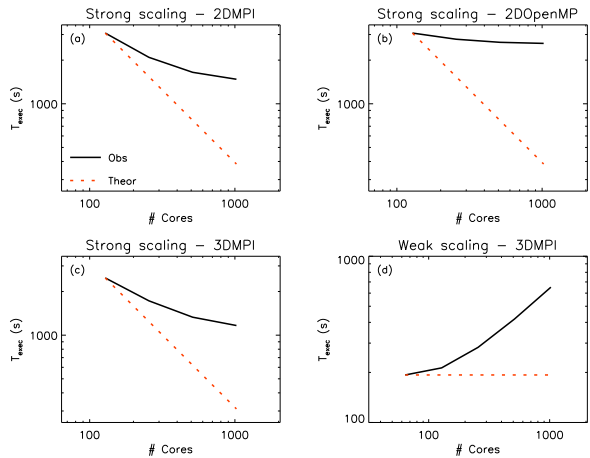}
	\end{center}
	\caption{(a) Strong scaling from 128 to 1024 cores on Marconi KNL using 2 OpenMP threads per Task MPI on the 2.5D setup. (b) Strong scaling using 64 MPI task and from 2 to 16 OpenMP Threads per task on the 2.5D setup. (c) Strong scaling from 128 to 1024 cores on Marconi KNL using 2 OpenMP threads per Task MPI on the 3D setup. (d) Weak scaling from 64 to 1024 cores on Marconi KNL using 2 OpenMP threads per Task MPI on the 3D setup.}
	\label{fig:Scaling}
\end{figure*}

ViDA numerically integrates VD equations in a six-dimensional phase-space ($3D$--$3V$: $x, y, z, v_x, v_y, v_z$). Only the $3D$ physical space is parallelized using cubic cells (squared in the $2D$ configuration). For implementing these tests, we have chosen $51^3$ velocity gridpoints for each particle VDF (protons and electrons), which represent a typical value adopted in production runs, and we have performed about $100$ timesteps per test (note that changing the step number does not act on the code scalability). Note that two VDFs are advanced in time through the ViDA algorithm, this limiting the number of spatial grid points per single processor and hence increasing the number of communications. 

As a first step, we have analyzed the parallel performance in the $2D$--$3V$ configuration, adopting a physical-space grid with $1024\times1024$ points. This setup requires about 6 TB of RAM, corresponding to, at least, 64 Marconi-KNL nodes. We have performed a strong scaling test by reducing the number of MPI Tasks per node from 8 to 1 and maintaining the same number of two OpenMP threads per task. Results are presented in Fig.~\ref{fig:Scaling}(a): the parallel efficiency scales efficiently up to 512 cores. As the number of cores increases, the efficiency is degraded owing to the more significant weight of MPI communications. This is mainly due to the huge memory request of the code combined with the Marconi KNL architecture. The code performance would strongly benefit from using a computer architecture with a larger RAM and a lower number of cores per node. We have also verified that the performance degradation cannot be handled by using an OpenMP strategy, as shown in  Fig.~\ref{fig:Scaling}(b), since the code performance is not affected by increasing the number of threads per node. In summary, within the current parallelization, the best performance is achieved with 32 MPI threads and 2 OpenMP tasks per node on a KNL system.

A slightly better performance is achieved using a full $3D$ configuration with $128\times128\times64$ grid points in the spatial domain. The strong scaling from $128$ to $1024$ cores is shown in Fig.~\ref{fig:Scaling}(c). A weak scaling test has been also performed by multiplying the number of spatial points and the number of cores (nodes) by the same factor. From the results, presented in Fig \ref{fig:Scaling}(d), it can be appreciated that the parallel efficiency is high only up to several hundreds cores. 

These preliminary tests show a reasonable parallel efficiency on KNL architecture, at least up to some hundreds cores. We are presently working to increase the code efficiency, in particular optimizing the communication pattern of the ViDA algorithm. We note that these results in part depend on the employed architecture. It is worth to finally highlight that, for instance, by using a Skylake machine with 192 GB and 48 cores per node, we would be able to increase by a factor of 3 the number of spatial gridpoints per node, thus increasing the parallel efficiency of the code, as the number of communications would strongly decrease. 

Concerning the computational costs, the ViDA code is about twice as computationally expensive as the HVM code \citep{valentini2007hybrid},  which has been recently used for 3D simulations of plasma turbulence (see for instance \citet{cerri2018dual}). More specifically, the reconnection run presented here -- which is the most expensive test  in this paper in terms of required computational resources --  has a cost of slightly less than 0.1 Mh on Marconi supercomputer using 16 nodes and 512 MPI processes. On the other hand, being ViDA a code for a new piece of physics, it is difficult to foresee for the exact cost of a 3D reconnection (or turbulence) run because the numerical and physical parameters, as well as the duration of the run, can vary significantly with respect to the standard ones used with the HVM code. Based on the experience with the HVM code, we may suggest that a high resolution 3D run of magnetic reconnection focusing on the electron physics would take from a few to a few tens of Mh. Such a significant allocation of computing time can be obtained, for example, in the framework of a PRACE project.

for a new piece of physics, it is difficult to foresee for the exact cost of a 3D reconnection (or turbulence) run because the numerical and physical parameters, as well as the duration of the run, can vary significantly with respect to the standard ones used with the HVM code.
\section{Conclusion}
\label{sect:concl}

In this paper we have presented a fully-kinetic code (ViDA) based on a Vlasov-Darwin algorithm, where only light waves are excluded in order to relax the constraint on the timestep advancement. This approach is particularly suited for the investigation of the kinetic dynamics from sub-ion scales down to the electron kinetic scales $d_e$ and to the Debye length $\lambda_{D}$. As typically the case for space plasmas, but often also in the laboratory, inter-particle collisions are not described, since collisional scales are assumed to be smaller than other characteristics dynamical scales.

ViDA has been tested against several waves modes, in particular Alfv\'en, whistlers and plasma waves. The development of the Weibel instability and reconnection, both in a regime where the main dynamics is driven by the electrons, has been also reproduced. These tests represent typical regimes of interest for studying the electron scale kinetic dynamics representing at today a strong computational challenge and a frontier problem for the understanding of the electron plasma physics. 

One of the main future objectives of ViDA will be the study of the structure and dynamics of the electron diffusion region, including the role of anomalous resistivity in the Ohm’s law and the mechanisms of electron heating, which are among the main targets of satellite MMS data analysis \citep{torbert2016estimates, genestreti2018mms, cozzani2019insitu}. Last but not least, we will make use of the ViDA code for the study of the plasma turbulent dynamics focusing on the problem of the "dissipative" scale, of primary interest in the context of the solar wind turbulent heating at kinetic scales \citep{vaivads2016turbulence}.

\acknowledgments
This project (FC, AR, FV) has received funding from the European Union’s Horizon 2020 research and innovation programme under grant agreement No 776262 (AIDA, www.aida-space.eu). OP is partly supported by the International Space Science Institute (ISSI) in the framework of the International Team 405 entitled ``Current Sheets, Turbulence, Structures and Particle Acceleration in the Heliosphere''. EC is partially supported by NWO Vidi grant 639.072.716. Numerical simulations discussed here have been performed on the Marconi cluster at CINECA (Italy), under the ISCRA initiative (IsC53\_MRVDS and IsB16\_VDMMS) and under the INAF-CINECA initiative (INA17\_C2A16 and INA17\_C2A16b).


\end{document}